# Modeling Motility of Kinesin Dimer From Molecular Properties of Individual Monomers

Dagong Fan1, Wenwei Zheng1, Ruizheng Hou1, Fuli Li2, Zhisong Wang1*

1Institute of Modern Physics, and Applied Ion Beam Physics Laboratory, Fudan University, Shanghai 200433, China

2College of Science, Xi'an Jiaotong University, Xi'an 710049, China

*Corresponding author (E-mail: wangzs@fudan.edu.cn)

**Abstract**

Conventional kinesin is a homodimeric motor protein that unidirectionally transports organelles along filamentous microtubule (MT) by hydrolyzing ATP molecules. There remain two central questions in biophysical study of kinesin: (1) the molecular physical mechanism by which the kinesin dimer, made of two sequentially identical monomers, selects a unique direction (MT plus end) for long-range transport; (2) the detailed mechanisms by which local molecular properties of individual monomers affect the motility properties of the dimer motor as a whole. On the basis of a previously proposed molecular physical model for kinesin's unidirectionality, this study investigates the dimer's synergic motor performance from well-defined molecular properties of individual monomers. During cargo transportation, and also in single-molecule mechanical measurements, a load is often applied to the coiled coil dimerization domain linking the two motor domains ("heads"). In this study the share of load directly born by each head is calculated, allowing an unambiguous estimation of load effects on individual heads' ATP turnover and random diffusion. The results show that the load-modulations of ATP turnover and head diffusion are both essential in determining the dimer's performance under loads. It is found that the dimer's consecutive run length depends critically on a few pathways leading to individual heads' detachment from MT. Modifying rates for these detachment pathways changes the run length but not the dimer's velocity, in consistence with mutants experiments. The run length may increase with ATP concentration or not, depending on a single rate for pure mechanical detachment. This finding provides an explanation to a previous controversy concerning ATP dependence of the run length, and related quantitative predictions of this study can be tested by future experiment. This study also finds that the experimental observations for assisting loads can be quantitatively explained by load-biased head diffusion. We thus conclude that the dimer motility under resisting as well as assisting loads is governed by essentially same mechanisms.

**Introduction**

Conventional kinesin is a homodimeric motor protein that transports vesicular cargos along filamentous cytoskeletal microtubule (MT) towards the MT plus end(*1*). Single-molecule experiments found that a kinesin dimer maintains its directionality against an opposing force of several picoNewtons(*2*). A kinesin dimer can walk tens or even hundreds of consecutive steps along before dissociating(*3*). This unique feature of long processivity empowers kinesin to serve as a long-distance transporter in intracellular vesicular traffic(*4*).

Conventional kinesin consists of a pair of globular motor domains ("heads"), which contain both microtubule- and nucleotide-binding sites. The two motor heads are connected by their neck linkers, which are each a ~14 amino acid peptide. The two neck linkers converge into a long coiled-coil that further connects the cargo-binding domains. A neck linker is in a random coil conformation when a MT-bound head is in nucleotide-free or ATP-bound state. Upon ATP binding at the MT-bound head, however, part of the neck linker adopts an ordered conformation and is immobilized to the catalytic core domain of the motor head(*5*). This conformational change, termed neck linker zippering, points the neck linker towards the MT plus end. Vale et al. suggested that neck linker zippering is the molecular mechanism for kinesin's plus-end directionality(*1, 5*).

Kinesin's processivity apparently requires a head-head coordination beyond the local event of linker zippering at individual heads. On one hand, Hancock and Howard(*6*) found that individual single-headed kinesin is not processive. On the other, Tomishige, Klopfenstein and Vale(*7*) found that a processive motor can be made by dimerizing Unc104/KIF1A that is a class of monomeric kinesin motors. Kinetic studies(*8-10*) suggested that kinesin's two heads alternately hydrolyze ATP. This kinetic mechanism of alternating head catalysis is consistent with the hand-over-hand walking gait of kinesin dimers found in single-molecule fluorescence measurements(*11*). Moreover, Thorn, Ubersax and Vale(*12*) found that the electrostatic interaction between the neck coiled coil and MT is a major factor affecting processivity. By engineering extra charged groups into the coiled coil, they created mutants that have a similar velocity but four-fold longer run length than wild-type kinesin. These results imply that the molecular factors responsible for processivity are largely uncoupled with those affecting velocity.

Controversies remain on kinesin's unidirectionality and processivity. First, mechanical experiments by Block et al.(*2, 3*) found that kinesin's processivity and velocity both depend on ATP concentration. In this experiment, a finite opposing force ≥ 1 pN was applied to the coiled coil domain using an optical clamp apparatus. However, a fluorescence experiment under zero load by Yajima et al.(*13*) found that kinesin's run length is independent of ATP concentration. Second, the neck linker zippering has a free-energy gain of merely ~ 1.2 $k_B T$ ($k_B$ is Boltzmann constant and $T$ is absolute temperature)(*14*). Several groups(*15, 16*) pointed out that the low zippering energy alone appears insufficient to account for the robust directionality of kinesin against loads of several picoNewtons. Although a large number of mechanical, kinetic and structural studies on monomeric and dimeric kinesins from drosophila, rat and human have been done, an unambiguous consensus has yet to be reached on molecular mechanisms underlying kinesin's unidirectionality and processivity. This fact is highlighted by the recent experiments of Cross et al.(*16, 17*) which suggest existence of a previously unknown "tethered state".

Previous theoretical studies on kinesin include thermal ratchet models, in which the kinesin dimer is treated as a single or two linked Brownian particles moving in periodic asymmetric potentials(*18-20*), discrete stochastic models(*21, 22*), and kinetic fitting models(*3, 23*). In several latest studies(*20, 24, 25*), more molecular details were considered. Recently, we proposed from physical first-principles a molecular mechanism(*26*) by which a dimer of individually directionless and nonprocessive heads selects the unique direction of MT plus end, and attains the synergic capability of processive run. This physical mechanism is termed molecular ratchet-and-pawl (MRP) mechanism. The MRP mechanism naturally gives rise to autonomous coordination between the two sequentially identical heads, and thereby the hand-over-hand walking gait. The MRP mechanism virtually prevents consecutive backward steps when the load remains below the stalling force of 5 – 8 pN(*26*). Under superstall loads, however, the mechanism becomes partially defective, and rare consecutive backsteps can occur(*27*). The predictions for kinesin's load-resisting capacity are in consistence with mechanical measurements(*2, 16, 28-30*). Interestingly, inker zippering and the length of the linker peptides are equally essential in kinesin's ratchet-and-pawl mechanism. For a linker length much larger than the wild type value, head-head coordination is lost, and linker zippering becomes insufficient to cause the robust plus end directionality. Another notable feature of the ratchet-and-pawl mechanism is that its working depends weakly on the zippering energy. Consequently, the low measured value is sufficient to reproduce a dimer's load-resisting capacity.

The MRP mechanism is a synergic physical mechanism, by which a homodimeric motor protein attains directionality and autonomous head-head coordination from first principles. Therefore, the MRP mechanism also offers concrete guidelines for detailed designs of kinesin-mimicking artificial nanomotors(*31*). The MRP mechanism provides a unified physics basis to study how molecular properties or processes of individual monomers affect the ultimate performance of the entire dimer. In this study we draw on the strength of the ratchet-and-pawl mechanism to develop a quantitative kinesin model that integrates molecular details with predicting power. This study has a threefold purpose. First, we shall focus on kinesin's processivity, and elucidate the molecular factors determining this synergic dimer-level motility property. The results will provide an explanation to the ATP-dependence controversy mentioned above. Kinesin's processivity directly bears biological functions in long-distance vesicle transportation, particularly in elongated nerve cell axons. Processivity is also a quantity related to thermodynamics of molecular motors in general. A track-walking motor is able to do useful mechanical work only when it runs persistently into a certain direction. No continual runs, no useful work. So a finite processivity imposes a limit on a motor's maximum output of useful work. An understanding of kinesin's processivity and the major molecular factors behind it will offer some valuable guidelines on development of kinesin-mimicking nanomotors(*31-33*). Second, we shall quantitatively examine the effects of head diffusion on kinesin's motility. This physical process can be estimated to a fair extent of certainty using the established method of the first-passage time theory(*32, 34, 35*). A head's diffusion is affected by the neck linkers connecting the two heads, is also affected by loads added to the coiled coil dimerization domain, and by viscosity and temperature of the solution environment. Third, we shall investigate the load-dependence of ATP turnover rates of individual heads, which is a central issue in kinesin study. The load dependences of ATP turnover and of head diffusion compete to determine the dimer-level motor performance. With a relatively easy estimation of the latter effect, we will be able to examine the former effect unambiguously.

## MODEL AND METHOD

### Molecular ratchet-and-pawl mechanism

The ratchet-and-pawl mechanism, as identified in our previous study(*26*), selects and locks a kinesin dimer's movement into the MT plus-end. This molecular mechanism for kinesin's unidirectionality has been derived from a molecular mechanical calculation for all possible dimer-MT binding configurations. As found by the calculation, concomitant binding of the two heads with MT stretches the neck linkers so much that the elevated free energy of the linker peptides becomes comparable to head-MT binding energies (we note that double-headed dimmer-MT binding configurations have been directly observed in a recent experiment(*36*)). As a consequence, a unique hierarchy for dimer-MT configurations occurs in terms of their total free energies. Namely, the double-headed dimer-MT configuration in which only the rear head is accompanied by linker zippering (configuration 1 illustrated in Fig. 1 B) has the lowest free energy, while any double-headed configurations with linker zippering at the front head are extremely high in energy and are virtually inaccessible. This configurational hierarchy imposes a restriction on the two heads. When both heads are MT-bound, ATP binding to the rear head occurs readily because the consequent linker zippering lowers the overall configurational energy. On the contrary, ATP binding to the front head is prohibited because the ensuing zippering amounts to a transition to one of the inaccessible configurations. Thus, kinesin's configurational hierarchy provides a mechanical basis for the kinetic mechanism of alternating head catalysis, which in turn guarantees kinesin's processivity. More importantly, the configurational hierarchy yields a translationally asymmetric lowest-energy configuration that occurs most frequently according to Boltzmann's law and thus dominates kinesin's interacting dynamics with MT. The lowest-energy configuration is asymmetric in the sense that neck linker zippering is allowed at the rear head but not at the front head. The asymmetric lowest-energy configuration is the basis for kinesin's directionality.

When kinesin binds to MT in the lowest-energy configuration, ATP hydrolysis readily occurs at the rear head to initiate the head detachment. But a similar ATP-initiated detachment is prevented at the front head. During the diffusion of the detached head, the former front head is allowed to bind ATP. The ensuing zippering at the MT-bound head will bias the mobile head's diffusion towards the MT plus-end. If the diffusing head binds successfully to a forward site, a forward step occurs and the dimer resumes the lowest-energy configuration. As schematically illustrated in Fig. 1 B, the mechanochemical cycle for forward stepping is essentially the same as proposed by Vale and Milligan(*1*). If the diffusing head otherwise binds back to its original position, which may occur if the MT-bound head remains in nucleotide-free state, a futile step occurs. In this case the dimer resumes the lowest-energy configuration too. Again it is the rear head but not the front head that will get ready to bind ATP and detach for another attempt for forward stepping. In summary, the kinesin dimer always moves towards the MT plus-end as if the movement were locked into this direction. Thus, the plus-end directionality of kinesin dimer arises from a synergic mechanism that involves a unique energy hierarchy for binding configurations of the entire dimer with MT. When the dimer is truncated into individual monomers, the synergy is destroyed and the robust directionality is lost.

Thus, kinesin is essentially a molecular ratchet-and-pawl device, in which the asymmetric lowest-energy configuration serves the role of "ratchet" and the selective detachment of the rear head the role of "pawl".

**Kinetic Monte Carlo simulation**

In this study, we conducted a kinetic Monte Carlo simulation for kinesin's walking dynamics within the framework of the ratchet-and-pawl mechanism. For details of the simulation method the reader is referred to Ref. (*26*). If the energy difference between two dimer-MT binding configurations is higher than the energy released from ATP hydrolysis, the transition between the two configurations is forbidden. This energy requirement naturally incorporates the transition rules derived from kinesin's configurational hierarchy. Rates used in the simulation are indicated in Fig. 1 C.

The simulation used experimental values for the enzymatic rates of individual heads. The following rates were taken from ref. (*37*) and references therein: ATP binding rate $k_1$ = 3.2 µM s$^{-1}$, reverse dissociation rate $k_{-1}$ = 150 s$^{-1}$, hydrolysis rate $k_2$ = 180 s$^{-1}$, rate for reverse ATP synthesis $k_{-2}$ = 18 s$^{-1}$, rate for γ-phosphate release $k_3$ = 250 s$^{-1}$, rate for ADP release from a MT-bound head $k_4$ = 300 s$^{-1}$. ADP release from a detached head is extremely slow and ignored in this study. The chance for rebinding of hydrolysis products (phosphate and ADP) is small because of their low concentrations (<10 nM) in the motility assay(*3*). So the steps of product release were assumed irreversible.

As found by experiments(*38-40*), a head binds to MT strongly in ATP or ADP·Pi state, but weakly in ADP state. For double-binding configurations, detachment of the ADP-associated head is likely assisted by the mechanical strain of the neck linkers(*1*). Thus, we assumed in the simulation that release of γ-phosphate from the catalytic core within a head triggers detachment of the ADP-bearing head from any double-headed dimer-MT configuration. Detachment of an ADP-bound head from a single-headed dimer-MT configuration was treated differently (see following subsections).

**Pathways for derailment of the entire kinesin dimer from MT**

The kinesin dimer derails entirely from MT when both heads are detached. The major pathways leading to derailment are single-headed dimer-MT binding configurations, from which detachment of the single MT-bound head derails the entire dimer. As illustrated in Fig. 1 B, three single-headed configurations occur during kinesin's movement, with the MT-bound head respectively in nucleotide-free state (marked configuration 2), ATP/ADP·Pi-bound state (configuration 3) and ADP-bound state (configuration 4). The rates for head detachment from these single-headed dimer-MT configurations, $k_5$, $k_6$ and $k_7$ have been previously measured by several groups(*10, 37, 40-42*). These measured values often differ from experiment to experiment. In this study, we obtained the values for $k_5$, $k_6$ and $k_7$ by fitting the simulation results to the experimental data of velocity and processivity(*2, 3*). The rates from experiments and from our fitting are presented in Table 1. As can be seen, the values from the fitting are well within the range spanned by experimental values.

In the single-headed dimer-MT configurations, the MT-bound head alone bears the load. The values of $k_5$, $k_6$ and $k_7$ should in principle change with changing load. Assuming a Arrhenius-type load-rate relationship $k = k(f=0)\exp[f \times d]$, we did trial calculations to fit the measured velocity and processivity. The results show that reasonable fit requires a small value for $d$. For example, the $d$ value for $k_5$ must be as small as ~ 0.2 nm. This finding agrees with a recent experiment(*15*), which found only weak dependence of the overall derailment rate on load in the pre-stall range, particularly at room temperature. We therefore neglected the load dependence of $k_5$, $k_6$ and $k_7$, because this study focuses on kinesin's stepping under loads below the stall force.

The three single-headed dimer-MT configurations as derailment pathways have different ATP dependence. Derailment from the configuration with the MT-bound head free of nucleotide may be termed pure mechanical derailment, while derailment from the other two configurations with the MT-bound head carrying post-hydrolysis ADP or ATP are both nucleotide-dependent, and may be termed as post-hydrolysis and ATP-accompanied derailment, respectively. The total probability for derailment is a sum of contributions from each derailment pathway, i.e.

$$p_{derail} = p_{mec} \cdot t_E \cdot k_5 + p_{APT} \cdot t_T \cdot k_6 + p_{ADP} \cdot t_D \cdot k_7 \ . \qquad (1)$$

Here $p_{mec}$, $p_{ATP}$ and $p_{ADP}$ are the occurring probabilities of the single-headed dimer-MT binding configurations leading to mechanical, ATP-accompanied and post-hydrolysis derailment, respectively. $t_E$, $t_T$ and $t_D$ are durations of the corresponding configurations. Considering the processes leading out of these configurations (see illustration Fig. 1), and neglecting the small values of detachment (Table 1), we can roughly estimate the durations as

$$t_E \approx \frac{1}{k_1[ATP]+k_{b1}+k_{f1}}, \qquad (2)$$

$$t_T \approx \frac{1}{k_2+k_{b2}+k_{f2}}, \qquad (3)$$

$$t_D \approx \frac{1}{k_4+k_{b1}+k_{f1}}. \qquad (4)$$

**Barriers and forces**

During kinesin's steps a mobile head reaches a nearby binding site on MT via intra-chain diffusion in which the linkers are self-stretched. The self-stretching drains conformational entropy out of the linker chains, and causes a free-energy barrier for kinesin's steps. Linker zippering at the MT-bound head points the diffusing head towards the binding site to the MT plus end, thus reducing the barrier for forward steps(*5, 14*). Therefore, the lowest barrier for forward stepping occurs when the MT-bound head is in ATP or ADP·Pi state, and the lowest barrier for backward stepping occurs when the MT-bound head is nucleotide-free (see illustration in Fig. 1 B). The barriers for forward and backward stepping were obtained by calculating the Helmholtz free energy of the linker peptides for the dimer-MT configurations involved. For zero load, the calculation yielded a barrier difference between forward and backward steps as ~ 6 $k_B T$, in fair agreement with the experimental value(*15*). This explains how the small zippering energy (~ 1.2 $k_B T$) can be amplified into a much larger diffusional bias.

The Helmholtz free energy of the neck linkers was calculated using a worm-like chain formula that has been verified by mechanical measurements on single polymers. The persistence length for kinesin's neck linkers was determined on the basis of an atomic computation for the linker peptides.
For details of the linker peptide calculation, the reader is referred to our previous publication(*26*).

As a common practice for mechanical experiments on kinesin, a force is applied to the neck coiled coils (see illustration in Fig. A). However, the diffusion and ATP turnover of an individual head is affected only by the amount of load it directly bears. In this study we calculated the forces upon individual heads in the following way. We first determined the extensions of both linkers by balancing forces at the coiled coil domain. The Gibbs free energy for the neck linkers was then calculated by combining the linkers' internal Helmholtz free energy with the contribution of the external force(*43*). In single-headed binding configurations the MT-bound standing head alone bears the load. In a double-headed binding configuration the neck linker adjacent to the front head is more extended than the linker adjacent to the rear head. Consequently the forces inflicted upon the two MT-bound heads by their adjacent neck linkers are different, and are given by derivatives of the Helmholtz free energies of the respective linker peptides. The calculated force for individual heads was used to consider load-dependence of kinesin's steps.

**Load-dependent diffusive search-and-bind rates**

The rate for a diffusing head to bind MT was calculated using the first passage time theory(*32, 34, 35*). The rate calculation considered the diffusional barriers as well as the geometry of the initial single-headed binding configuration and the final double-headed configuration(*26*). The head diffusion coefficient was taken as $D = 2.3 \times 10^6$ nm$^2$/s.

The calculated rate for forward diffusive binding assisted by zippering ($k_{f2}$ indicated in Fig. 1 B) and the calculated rate for backward binding accompanied by a nucleotide-free MT-bound head ($k_{b1}$) are presented in Fig. 5 E. With increasing load, the forward rate $k_{f2}$ decreases and the backward rate $k_{b1}$ increases, both exponentially. Also shown in Fig.5 E are the measured overall rates(*15*) for forward and backward steps, which place a lower limit for the diffusive binding rates. As can be seen from the figure, the calculated diffusive binding rates are close to the measured stepping rates at low loads, and deviate with increasing load.

**Load-dependent ATP turnover rates**

The enzymatic rates of a MT-bound motor domain are directly affected by the amount of force inflicted upon it by its adjacent linker peptide. In this study we adopted a realistic treatment of the enzymatic rates' load-dependence by calculating the forces directly born by individual heads in all possible dimer-MT binding configurations. It was assumed that the enzymatic rates of a MT-bound head are affected only if the force-transmitting neck linker points to MT minus end. Following the conformationally-composite state model of Schnitzer, Visscher and Block(*3*), we assumed that the ATP hydrolysis rate ($k_{hyd}$) and the ATP dissociation rate ($k_{off}$) depend on the rear-pointing force ($F_{head}$) by a Boltzmann-type relationship

$$k_{hyd}(F_{head}) = k_{hyd}(F_{head}=0)/[p_1 + q_1 \exp(F_{head}\delta/k_B T)], \qquad (5)$$

$$k_{off}(F_{head}) = k_{off}(F_{head}=0)/[p_2 + q_2 \exp(F_{head}\delta/k_B T)]. \qquad (6)$$

Here $p_1 + q_1 = p_2 + q_2 = 1$. We assumed $p_1 = p_2$, and used $q_1 = 0.0062$ as deduced by Schnitzer, Visscher and Block in Ref.(*3*). Through this study we used $\delta$ =2.7 nm. This value was suggested by Schnitzer, Visscher and Block in their recent experiment(*44*). We note that $F_{head}$ entering the above equations is the amount of force calculated for individual MT-bound heads. For a single-headed dimer-MT binding configuration, $F_{head}$ for the MT-bound head equals to the external load ($F$) that is applied to the coiled coil dimerization domain. For a double-headed kinesin-MT binding configuration, $F_{head}$ for either head does not equal to the external load.

**RESULTS**

**Velocity**

Fig. 2 A and B present kinesin's average velocity as a function of ATP concentration and load, respectively. The predictions of the present model agree fairly well with the experimental data of Visscher, Schnitzer and Block(*2*). We note that the agreement between theory and experiment is improved compared to our previous study(*26*).

This can be attributed to a slight adjustment of values for $k_1$ (ATP binding rate), $k_2$ (hydrolysis rate), $k_{-2}$ (rate for reverse ATP synthesis), and $D$ (head diffusion coefficient). Fig. 2 B also presents the predicted velocity for assisting load. At saturation ATP concentrations e.g 2 mM, the velocity is almost unchanged as the assisting load increases to 6 pN. At limiting concentrations e.g. 5 μM, however, the velocity slight increases with increasing assisting load. These features are consistent with the experiment of Block et al.(*44*). This study does not find any sudden change in the velocity-load curve around zero load, unlike the previous theory of Fisher and Kim(*24*).

Under negligible load, and at saturation ATP concentrations, hydrolysis and Pi release are rate-limiting processes for kinesin's stepping. Accordingly, kinesin's velocity is upper-limited by $d \times k_2 k_3 / (k_2 + k_3)$, where $d$ = 8.2 nm is the protofilament lattice spacing of MT. The rate values adopted by this study yield the upper limit as ~ 870 nm/s. As can be clearly seen in Fig. 2, kinesin is already close to this limit at ATP concentration of 2 mM. Similarly, ATP binding is the rate-limiting process for kinesin's stepping at low ATP concentrations and under low loads, and the corresponding limit on velocity is $d \times k_1 \times [ATP] \approx$ 130 nm/s for concentration $[ATP]$ = 5 μM. Kinesin is close to half of this limit, as can be seen in Fig. 2 B.

**Processivity**

In Fig. 3 A and B we compare the average consecutive run length predicted by this study to experimental data of Schnitzer, Visscher and Block(*3*). Again, the comparison shows a fair agreement between theory and experiment. The run length decreases with increasing load, largely due to the load-dependence of the rate for forward diffusive binding $k_{f2}$ (see Fig. 5 E). As the diffusive binding of a mobile head to MT becomes slower under higher loads, the time duration for single-headed dimer-MT binding configurations becomes longer (see eq. 2 - 4). Consequently, lower diffusive binding rates generally cause a higher derailment probability, and thereby a shortened run length.

Both this theory and the experiment of Schnitzer, Visscher and Block(*3*) found that the run length first increases as the ATP concentration is raised from limiting values, and remains flat as the concentration is further raised beyond a few tens μM. This ATP-dependent processivity can be understood by analyzing ATP-dependence of the derailment pathways. In Fig. 3 C we present the normalized percentage of derailment events through different derailment pathways as a function of changing ATP concentration for a low load of 1.05 pN. It is clear from the figure that the pure mechanical derailment is dominant for $[ATP]$ < 3 μM, but its role quickly diminishes as the concentration is raised. For the low concentrations $[ATP]$ < 3 μM, the total derailment probability can be roughly estimated using eq. 1 and 2,

$$p_{derail} \approx p_{mec} \cdot k_5 \cdot \left( \frac{1}{k_1[ATP] + k_{b1} + k_{f1}} \right). \quad (7)$$

Here $p_{mec}$ can be reasonably regarded as being proportional to the percentage given in Fig. 3 C. As a crude approximation, we have $p_{mec} \propto \frac{1}{[ATP]}$. Therefore, the derailment probability increases with decreasing ATP concentration with a double ATP dependence. This sharp ATP dependence is the reason for the decrease of the run length with dropping concentration for the range $[ATP]$ < 3 μM.

As the ATP concentration is raised from 1 to 20 μM, the percentages of ATP-accompanied and of post-hydrolysis derailment both increase (see Fig. 3 C). Again, the percentages can be reasonably regarded as being proportional to $p_{ATP}$ and $P_{ADP}$ in eq. 1. Because $t_T$ and $t_D$ are both ATP independent (see eq. 3 and 4), the ATP dependence of the contributions from the ATP-accompanied and post-hydrolysis pathways to the total derailment probability is solely determined by $p_{ATP}$ and $P_{ADP}$. Consequently, the portion of derailment probability caused by the two pathways should decrease with decreasing ATP concentration over the range of 1 to 20 μM. Therefore, the two pathways generally cause an opposite ATP dependence of the run length compared to the pure mechanical pathway at low ATP concentrations.

For $[ATP]$ > 20 μM, the percentages for the ATP-accompanied derailment and for the post-hydrolysis derailment both become flat as can be seen in Fig. 3 C. Over this high concentration range the pure mechanical pathway is comparatively negligible. Then the total derailment probability can be approximated as $p_{derail} \approx p_{ATP} \cdot t_T \cdot k_6 + p_{ADP} \cdot t_D \cdot k_7$. Noting again the ATP independence of $t_T$ and $t_D$, the flattening of the derailment event percentages explains the flattening of the run length over the concentration range beyond 20 μM.

The results in Fig. 3 C further show that for the broad range of concentrations beyond $[ATP]$ = 10 μM, the post-hydrolysis derailment is dominant, in agreement with the experimental finding of Thorn, Ubersax, and Vale(*12*).

In a brief summary, the pure mechanical pathway is the dominant factor behind the ATP dependence of the run length for low ATP concentrations. The post-hydrolysis and ATP-accompanied pathways determine a virtually ATP-independent run length for high concentrations, with the former as the major determinant of the magnitude of the run length.

**A possible solution to the controversy of ATP-dependence of processivity**

A long standing controversy concerning kinesin's processivity can be solved by an analysis of the three pathways' influence on processivity identified in this study. Using a force clamp apparatus on squid conventional kinesin, Schnitzer, Visscher and Block (*3*) found a Michaelis-Menten type dependence on ATP concentration, which has been well reproduced by the present theory (Fig. 3 A and B). A different set of experiment on rat conventional kinesin by Yajima et al.(*13*) using a fluorescence microscope found no obvious ATP dependence in a similar range of concentrations for zero load. As shown in Fig. 4, the present theory is able to reproduce satisfactorily the data of Yajima et al. on processivity, velocity and dwell times by merely adjusting two detachment rates. The head detachment rate related to the pure mechanical pathway ($k_5$) was reduced from 0.05 s$^{-1}$ to 0.03 s$^{-1}$, and the detachment rate related to the post-hydrolysis pathway ($k_7$) was raised from 25 s$^{-1}$ to 80 s$^{-1}$. As can be seen in the figure, the predicted run length remains flat as the ATP concentration drops from 2 mM down to 2 μM, in agreement with the data of Yajima et al.. The flattening of the run length-concentration curve at low concentrations can be attributed to the slight decrease of $k_5$, which reduces the doubly ATP-dependent contribution of the mechanical pathway to the total derailment probability. A larger $k_7$ value for the post-hydrolysis pathway reduces the absolute magnitude of the run length without changing its ATP dependence. Thus, the different ATP dependences observed in Schnitzer, Visscher and Block's experiment and in that of Yajima et al. are likely due to use of kinesin from different species. The conventional kinesin from squid and that from rat might possess different rates for the pure mechanical derailment and for the post-hydrolysis derailment. Future measurement of the two derailment rates (i.e. $k_5$ and $k_7$) for squid and rat kinesin will be able to test the above prediction of this study.

The scenario of a lower detachment rate for a nucleotide-free head as a solution to the processivity controversy was suggested before by Diez, Schief and Howard in an qualitative analysis(*45*). This study provides a quantitative basis for the scenario. As also shown in Fig. 4, the predicted run length decreases slightly as the concentration drops beyond 2 μM, i.e. beyond the range where the experimental data are available. If Yajima et al. would extend their measurement to lower concentrations, they might observe a weak ATP dependence. This possibility was also suggested before by Diez, Schief and Howard(*45*). We would like to note that the velocity data of Yajima et al. would be better fitted if a higher hydrolysis rate for rat kinesin (> 500 s$^{-1}$)(*42*) is adopted in this study.

**Load-dependence of ATP turnover**

Load-dependent catalytic rates for individual heads are a natural projection of rate theories, and were often assumed in fitting models(*3, 23*) for kinesin motility. By fitting a comprehensive set of experimental data on velocity and processivity, Schnitzer, Visscher and Block(*3*) obtained quantitative forms for load-dependent catalytic rates. The expressions used in fitting models are generally in terms of the load applied to the coiled coil dimerization domain. However, in a double-headed kinesin-MT binding configuration the external load is not the same as the forces directly born by individual heads. ATP turnover at an individual MT-bound head should be directly affected by the amount of force transmitted to the catalytic core within the head via the adjacent neck linker. As mentioned in the MODEL AND METHODS section, this study treated the load dependence of catalytic rates in terms of the linker-transmitted force instead of the total external load. This realistic treatment permits a conceptually sound estimation of the role of load-dependent ATP turnover in kinesin motility.

Load-dependent ATP turnover of MT-bound heads and load-sensitive diffusive binding of mobile heads are two major factors that compete to determine the overall load-dependence of kinesin motility. We have carried out two separate sets of simulations in which either the load-dependence of the catalytic rates or the load-dependence of diffusive binding rates was deliberately ignored. More specifically, in the first set of simulation we set $\delta$ = 0 in the Boltzmann-type relationship for catalytic rates (eq. 5 and 6); in the second set of simulations we adopted a fixed value of 500 s$^{-1}$ for the rate for zippering-assisted forward binding ($k_{f2}$). The results are presented in Fig. 5 A - D. It is clear that both sets of simulations severely overestimate kinesin's velocity at high loads. The discrepancies become bigger with increasing load. This finding suggests that the load-dependence of either ATP turnover or diffusive binding has indispensable effects on kinesin motility. The physical mechanism for load-dependence of diffusive binding is rather clear, and the first-passage time theory(*34*) used for calculating the corresponding rates is well established. Comparatively, the molecular mechanism for the load-dependence of ATP turnover remains unknown thus far. The finding that the simulations fully accounting the effects of the diffusive binding but ignoring the catalytic rates' load dependence are insufficient to explain the experimental data provides unambiguously a quantitative evidence for load susceptibility of ATP turnover in kinesin's steps.

**Effects of solution viscosity**

The viscous buffer solution environment in which kinesin molecules are measured experimentally has important effects on head diffusion. Generally, the rate for diffusive binding is proportional to the diffusion coefficient, which is in turn inversely proportional to the solution viscosity. To examine the influence of solution viscosity on kinesin motility, we conducted simulations using different diffusion coefficients for mobile heads' diffusive binding. Fig. 6 A presents the results for the velocity-load curve obtained using a diffusion coefficient 2 times and 0.66 times of the best-fit value (i.e. $D = 2.3 \times 10^6$ nm$^2$/s, which was used to obtain the results presented in Fig. 2 - 4). It can be seen in Fig. 6 A that a higher diffusion coefficient, corresponding to a less viscous buffer solution, tends to increase kinesin's velocity. This viscosity dependence of velocity is almost negligible at the limiting ATP concentration ~ 5 $\mu$M, but becomes notable when the concentration is raised to 2 mM.

Fig. 6 B presents the run length-concentration curves predicted using a diffusion coefficient only 20% higher than the best-fit value. The results show that a higher diffusion coefficient tends to improve processivity too. Processivity is the result of competition of the mobile head finding its next binding site with the MT-bound head detaching through various pathways. The faster the diffusing head searches and bind to MT, the better the processivity. Nevertheless, the change in diffusion coefficient does not modify the ATP dependence of the run length as can be seen in the figure. This finding rules out differing buffer solution viscosity as a possible explanation to the above mentioned controversy on ATP dependence of kinesin's processivity. The results in Fig. 6 A and B show that a change of diffusion coefficient by 20% can lead to apparent deviation from the best fits as in Fig. 2 to the experimental data of velocity and processivity. This finding implies that experiments using different buffer solutions may reasonably produce visibly differing data. This uncertainty should be considered in data analysis and in quantitative modeling.

**DISCUSSION**

**Detachment rates regulates processivity but with little influence on velocity**

In this study, we found that adjusting the rates for head detachment involved in derailment pathways changes the run length but not the velocity. This is easily understood because these derailment events terminate the run with but little influence on the occurred steps. These detachment rates are determined by head-MT affinity. Previous experiments(*38-40*) found that the head-MT binding is stronger for a nucleotide-free or ATP-bound head than for a post-hydrolysis ADP-bound head. A recent experiment found further that the binding between an empty head with MT is weakened right after ATP binding(*41*). Electrostatic interactions between the head and MT are largely responsible head-MT affinity. Thorn, Ubersax and Vale(*12*) engineered kinesin mutants with differing amount of electric charges in the neck stalk, and found more than tenfold difference in their run length but little difference in their velocity. This fact is readily understandable with findings of this study. A change in the electrostatic interactions between kinesin heads and MT likely modifies the detachment rates, particularly that for the ADP-carrying head weakly bound t MT. According to preceding analysis, the post-hydrolysis pathway for derailment is the dominant determinant for the run length at saturation ATP concentration. Our simulations show that increasing the detachment rate for this derailment pathway ($k_7$) by several times can elongate the run length by 10 times without changing much the velocity.

**Effect of backward diffusive binding**

In a single-head dimer-MT binding configuration, backward binding of the mobile head to MT is generally a low-rate process. The backbinding rate is ~ 1 s$^{-1}$ at zero load, and increases to ~ 10 s$^{-1}$ when the opposing force is raised to 6 pN (Fig. 5E). Furthermore, the mobile head's backward binding is virtually prohibited once the MT-bound head binds ATP, because the ensuing linker zippering at the MT-bound head raises the barrier for backward binding drastically (by several times of $k_B T$). Therefore, the relative magnitudes of the mobile head's backward binding rate and the MT-bound head's ATP binding rate determine the actual occurring probability of backbinding events. At saturating ATP concentrations, e.g. 2 mM, the ATP binding rate can be several thousands per second, making the backbinding virtually impossible. At limiting ATP concentrations, e.g. 5 μM, the ATP binding rate is as low as a few tens per second, which are of the same magnitude as the backbinding rate. Consequently, at limiting ATP concentrations the mobile head binds backwards to MT with a small but non-negligible probability. A backbinding event causes a futile step, thus more than ATP molecule will be consumed to make a successful forward step. Therefore, this model predicts that the average number of ATP molecules consumed for each forward step under low loads is one at saturating ATP concentrations but slightly above one at limiting concentrations(*26*). This conclusion is in consistence with a randomness measurement by Schnitzer and Block(*46*).

Under an assisting load, the backbinding event is suppressed because the load biases a diffusive head toward a forward site on MT. Extrapolating the exponential backbinding rate-load curve of Fig. 5 E to the assisting load regime, one finds that the backbinding rate is reduced by half by an assisting load of 4 pN as compared to the zero-load rate. This modest reduction in the backbinding rate, and thereby in the occurring probability of futile steps, explains the slight increase of the dimer velocity with increasing assisting loads for limiting ATP concentrations (Fig. 2B). At saturating concentrations, linker zippering readily occur to further reduce the backbinding events to a negligible level. Consequently, the velocity remains flat with increasing assisting loads at saturation concentrations. Thus, the load-biased diffusive binding quantitatively explains the major feature of kinesin motility under resisting loads. We thus conclude that kinesin's motility under resisting as well as assisting loads is governed by essentially same mechanisms.

**Conclusions**

In summary, we have investigated the detailed mechanisms by which multiple molecular properties of individual kinesin monomers determine the synergic motor performance of the dimer. The major conclusions are:

(1) An unambiguous estimation of load effects on individual heads' ATP turnover and random diffusion has been achieved by realistically calculating the share of load directly born by either of the two heads. The results show that the load-modulations of ATP turnover and head diffusion are both indispensable in determining the dimer's performance under loads.

(2) Processivity of the dimer motor, i.e. consecutive run length is a highly synergic property with regard to its apparent reliance on head-head coordination. Rather surprisingly, this study found that the dimer's processivity depends critically on three pathways leading to individual heads' detachment from MT. Changing rates for these detachment pathways changes the run length but not the dimer's velocity. This decoupling between run length and velocity is consistent with mutants experiments. The run length may increase with ATP concentration or not, depending on a single rate for pure mechanical detachment. This finding provides an explanation to a previous controversy concerning ATP dependence of the run length. The quantitative prediction related to the explanation can be tested by future experiment rather straightforwardly. Use of different buffer solutions with different values of viscosity affects the run length too, but does not change its trend of ATP dependence.

(3) The experimental observations for assisting loads can be quantitatively explained by considering the biasing effect of the load on a diffusive head. Load-biased diffusion occurs under assisting loads as well as resisting loads. The dimer's motility under either resisting or assisting loads is governed by essentially same mechanisms.

**Acknowledgement**

This work was partly funded by National Natural Science Foundation of China (Grant No. 90403006), Chinese Ministry of Education (under Program for New Century Excellent Talents in University), Shanghai Education Development Foundation (under Shuguang Program).

**Table 1. Rates for head detachment from a single-headed dimer-MT binding state (in s$^{-1}$).** The experimental values are from Hancock and Howard(*10*), Ishiwata et al.(*40*), Guydosh and Block(*41*), (*42*), and Cross(*37*). For comparison, hydrolysis-induced detachment of kinesin monomer is 65-90 s$^{-1}$ (ref.(*47*)).

| Rate | Hancock & Howard | Ishiwata et al. | Guydosh & Block | Auerbach & Johnson | Cross | This study |
|---|---|---|---|---|---|---|
| $k_5$ | 0.02 | 0.0067 | | | | 0.05 |
| $k_6$ | 0.01 | 0.0067 | 0.9 (3.9pN) | | | 0.5 |
| $k_7$ | 4 | 1 | | 9 | ~300 | 25 |

**Figure Captions**

**Figure 1**. (A) Schematic illustration of a kinesin dimer under a pulling force, which is often applied to the coiled coil dimerization domain in mechanical measurement. (B) Illustration of major dimer-MT binding configurations occurring in kinesin steps. The filled arrows indicate the mechanochemical cycle for a forward step, and the unfilled arrow indicates a possible pathway branching out of the main cycle. The three single-headed binding configurations, marked 2, 3 and 4 respectively and highlighted by rectangle frames, are major pathways for derailment of the entire dimer from MT. The mobile head in a single-headed configuration may bind to MT by random diffusion, and the rates for diffusive head binding are indicated. The circles and curves represent motor heads and neck linkers, respectively. The straightened part of the neck linker indicates a zippered state. The marks T, D and E indicate the ATP-bound state, ADP-bound state and nucleotide-free state of a head. The mark DP indicates a head to which the hydrolysis products ADP and Pi remain bound. (C) Mechanochemical states and rates of an individual head. The mark M·K indicates a MT-bound head (M stands for MT and K for kinesin

head), while K alone indicates a mobile head. E, T, D and DP indicate the head's nucleotide states as for (B). The solid arrows indicate transitions between the states occurring during kinesin's steps. The dotted arrows indicate the dimer's derailment.

**Figure 2.** Running speed of kinesin dimer. (A) Average velocity versus ATP concentration for various loads. (B) Average velocity versus applied loads for fixed ATP concentrations. The symbols are experimental data from ref.(*2*) for squid conventional kinesin, the lines are predictions of this study. The Major parameters used in this study are: rates for ATP binding and reverse dissociation $k_1$ = 3.2 µM$^{-1}$s$^{-1}$, $k_{-1}$ = 150 s$^{-1}$; rates for ATP hydrolysis and reverse re-synthesis $k_2$ = 180 s$^{-1}$, $k_{-2}$ = 18 s$^{-1}$; Pi release rate $k_3$ = 250 s$^{-1}$; ADP release rate $k_4$ = 300 s$^{-1}$ (for a MT-bound head). The rates for head detachment through the pure mechanical pathway, the ATP-accompanied mechanical pathway and the post-hydrolysis pathway are $k_5$ = 0.05 s$^{-1}$, $k_6$ = 0.5 s$^{-1}$, $k_7$ = 25 s$^{-1}$, respectively. The coefficient for head diffusion is $D$ = 2.3 × 10$^6$ nm$^2$/s.

**Figure 3**. Mean run length of kinesin dimer under finite loads. (A) Run length versus ATP concentration for various loads. (B) Run length versus applied loads for fixed ATP concentrations. The symbols are experimental data from ref.(*3*) for squid conventional kinesin. The lines are predictions of this study using the same parameters as form Fig. 2. (C) Percentage of dimer derailment from MT through different head detachment pathways predicted by this study for a small load of 1.05 pN.

**Figure 4**. Mean run length, velocity, and dwell time for kinesin's steps under zero load. The lines are predictions of this study using the head detachment rates $k_5$ = 0.03 s-1, $k_7$ = 80 s-1. All other parameters are same as for Fig. 3. The symbols are experimental data from Yajima and Cross(*13*) for rat conventional kinesin.

**Figure 5.** Effect of load-dependence of head diffusion and ATP turnover on kinesin's steps. (A, B, C and D) The lines are the results for kinesin's velocity from the calculations in which we deliberately ignored either the load-dependence of ATP turnover rates (A and B) or the load-dependence of the rates for diffusive binding of a mobile head (C and D). The symbols are experimental data from ref.(*2*). (E) Rates for a mobile head's diffusive binding to a site before or behind the other MT-bound head. The lines are predictions of this study. The symbols are the overall rates for forward and backward steps measured by Taniguchi et al.(*15*).

**Figure 6**. Effect of diffusion coefficient on kinesin's steps. The experimental data (symbols) are taken from refs.(*2*). The lines are predictions using the same parameters as for Fig. 2 and 3 except for values for the head diffusion coefficient. (A) Velocity versus load. The dashed and dotted lines are the results obtained using a diffusion coefficient 2 times and 0.66 times that for Fig. 2B. (B) Processivity versus ATP concentration. The dashed lines are the results obtained using a diffusion coefficient 1.2 times that for Fig. 3A.

# Figure 1

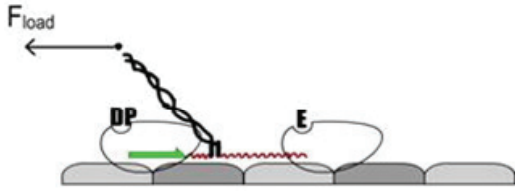

A

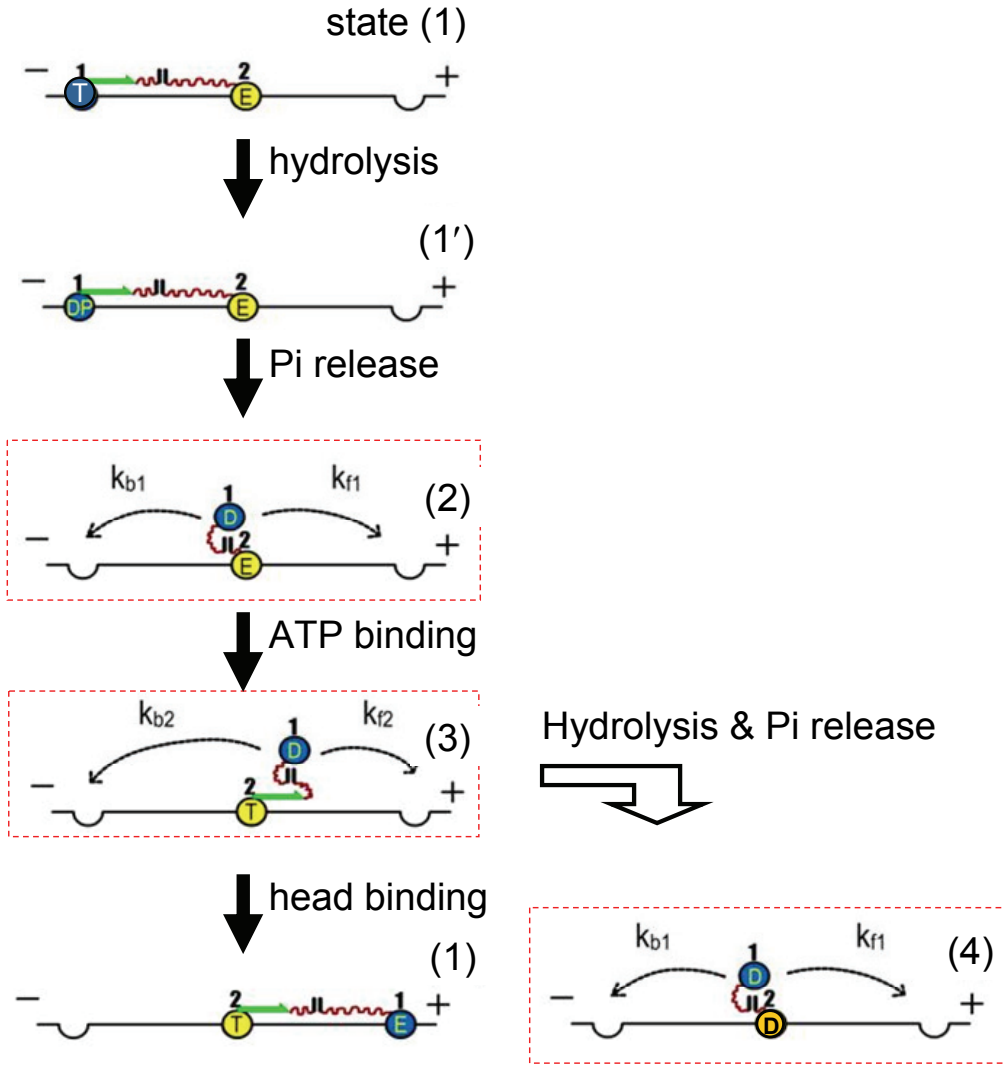

B

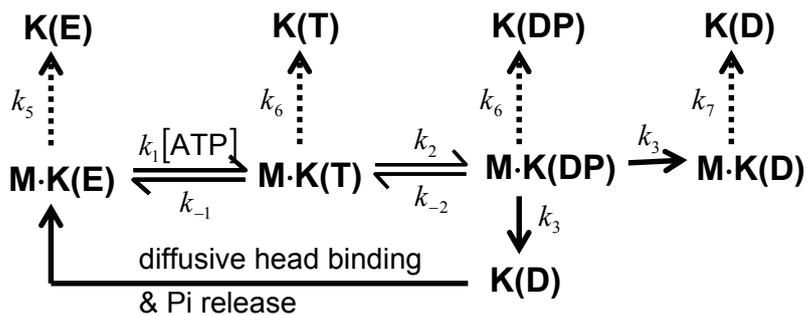

C

Figure 2

A

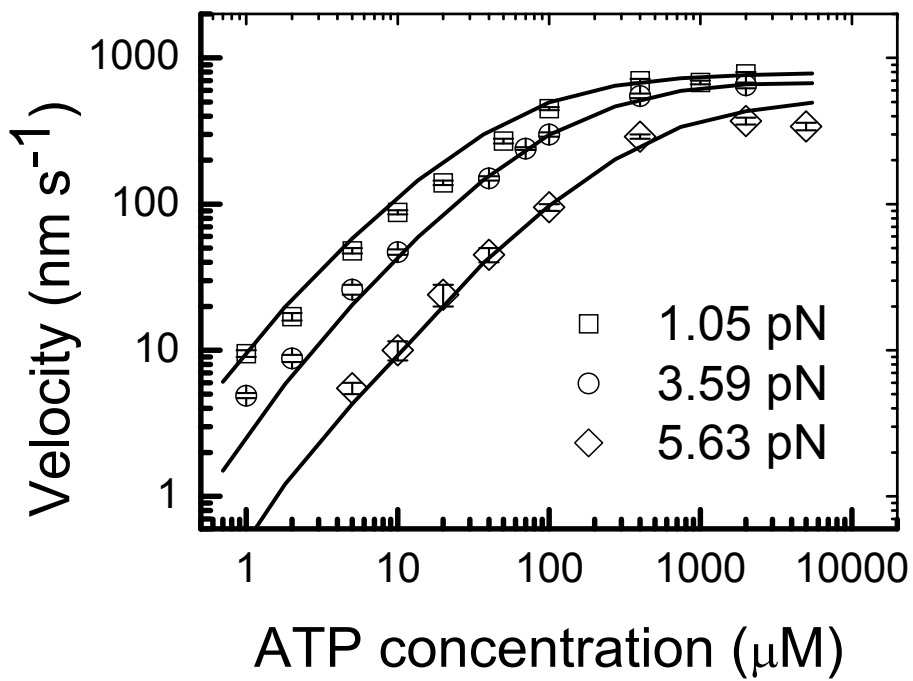

B

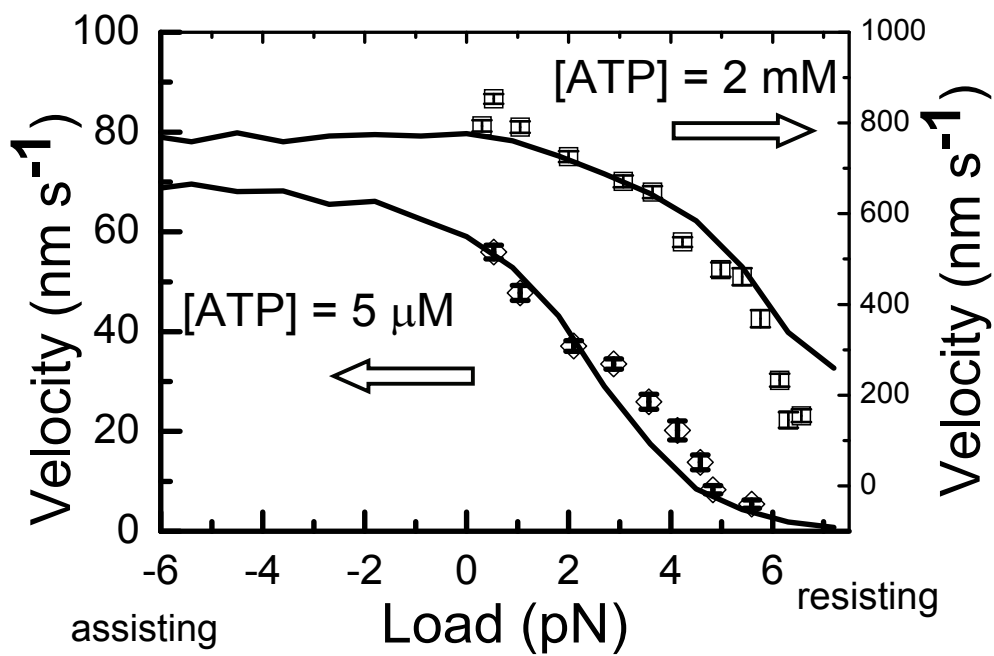

Figure 3

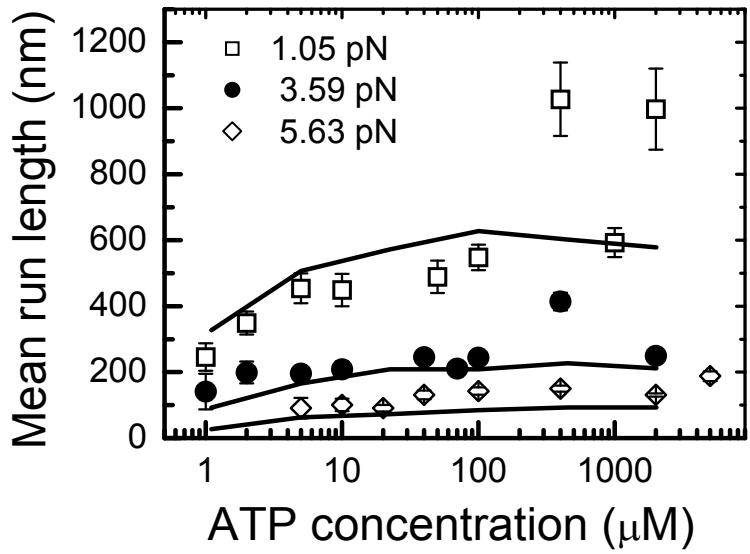

A

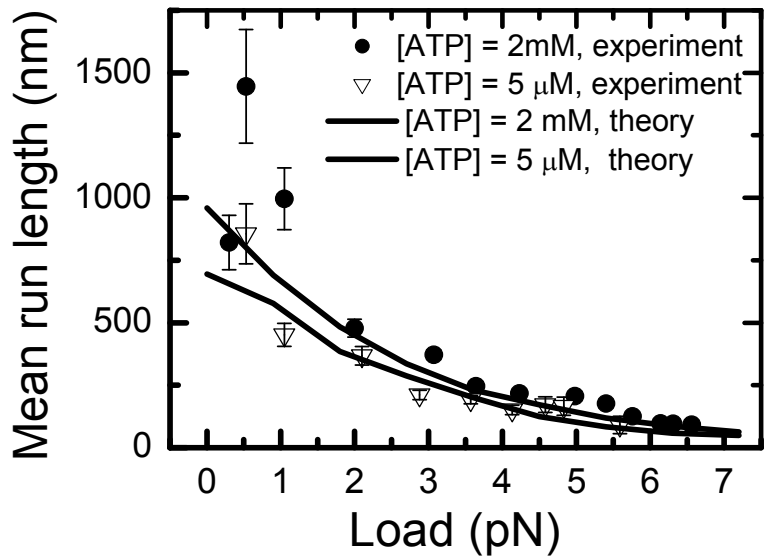

B

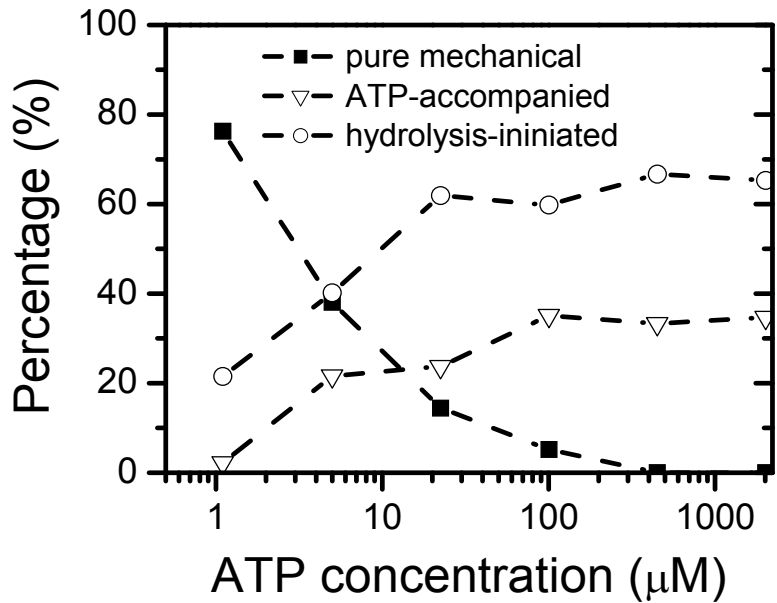

C

Figure 4

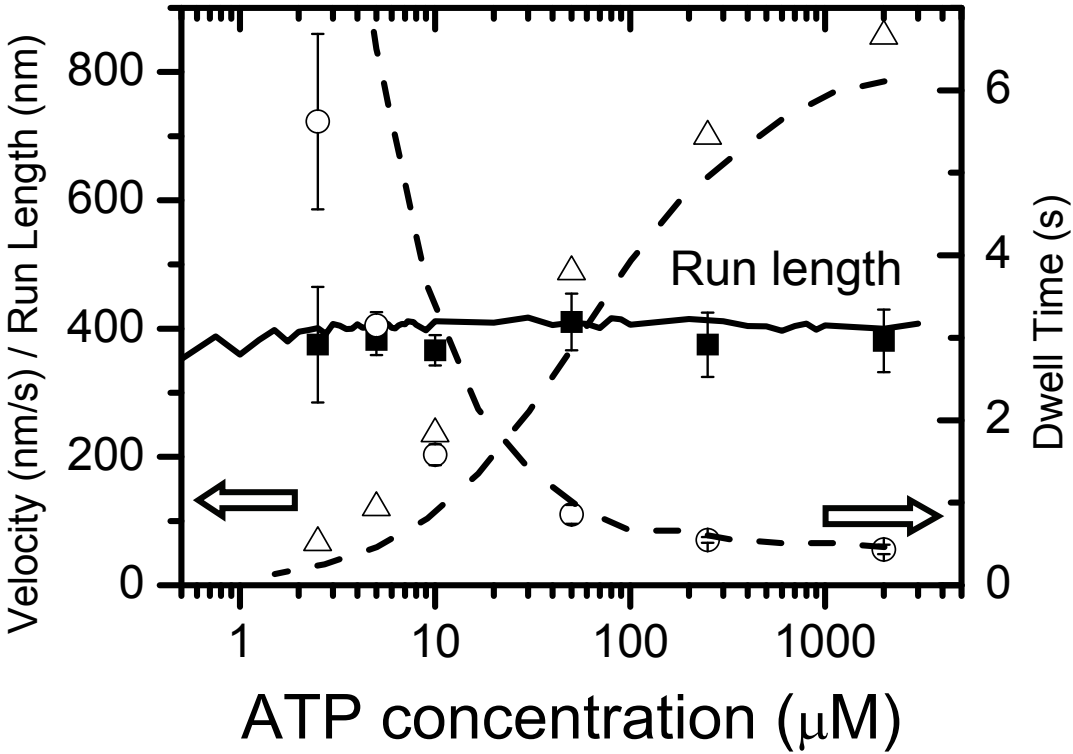

Figure 5

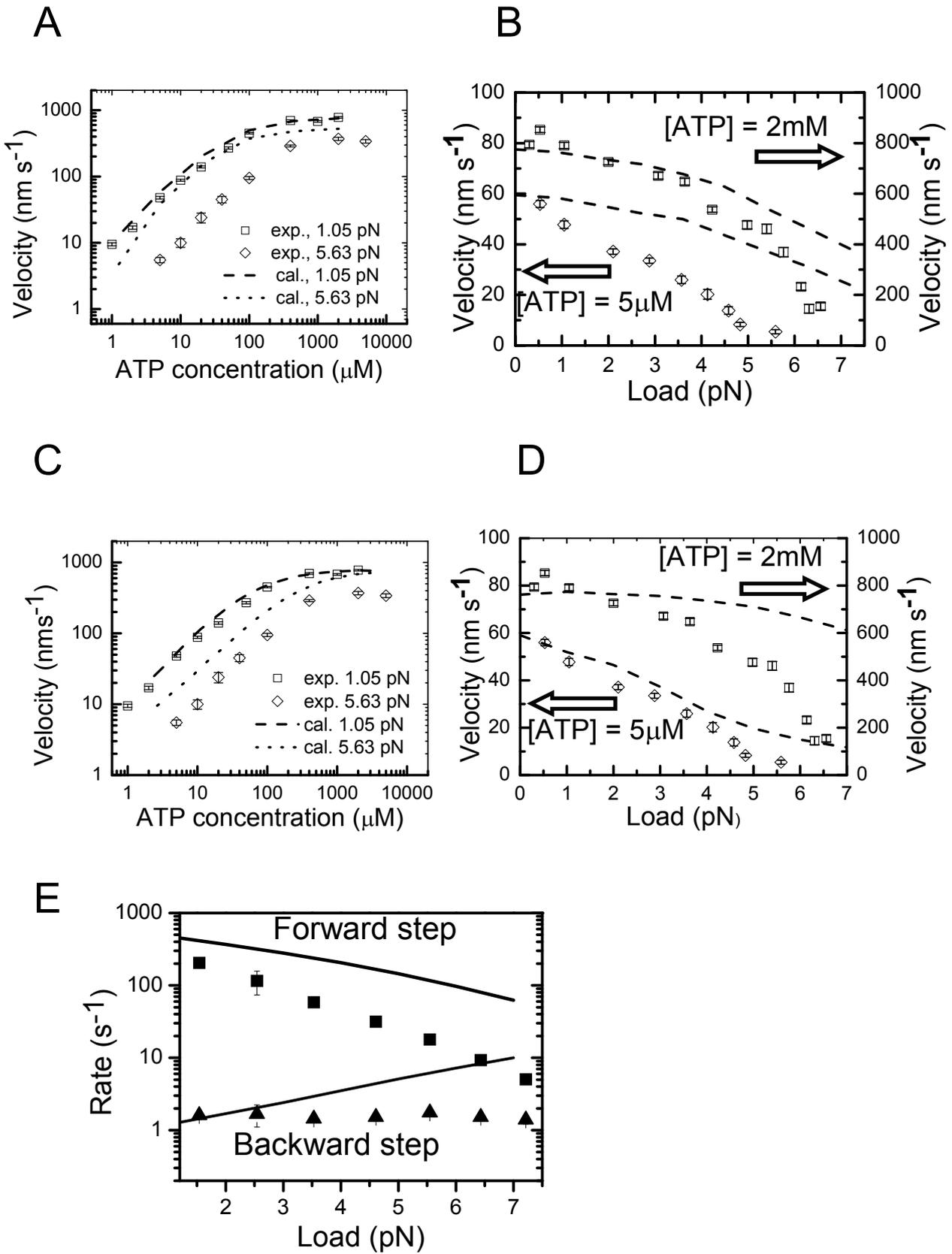

Figure 6

A

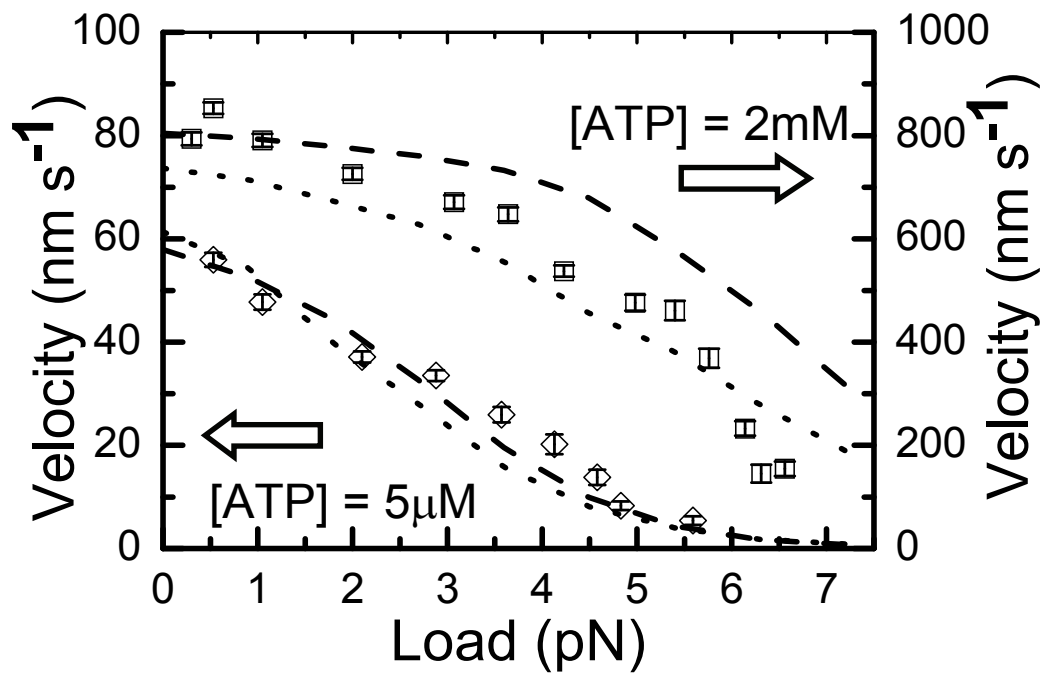

B

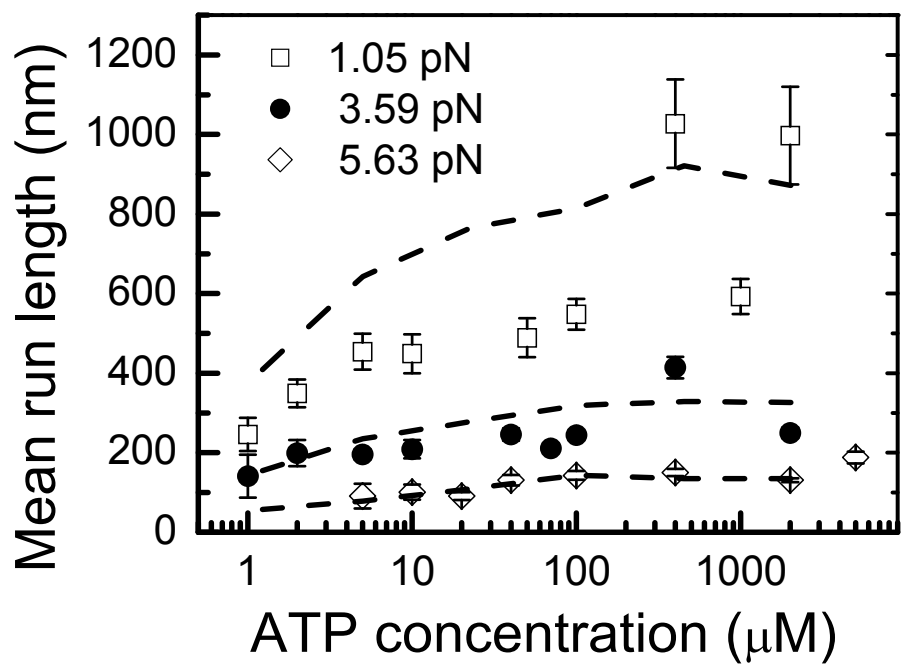